# Analysis of Bus Tracking System Using Gps on Smart Phones

[1]Mr. Pradip Suresh Mane, [2]Prof. Vaishali Khairnar
[1]PG Scholar,[2]Asst.Professor, Dept. of Information Technology, Terna Engineering College, Mumbai, India

***Abstract:*** *Public transport networks (PTNs) are difficult to use when the user is unfamiliar with the area they are traveling to. This is true for both infrequent users (including visitors) and regular users who need to travel to areas with which they are not acquainted. In these situations, adequate on-trip navigation information can substantially ease the use of public transportation and be the driving factor in motivating travellers to prefer it over other modes of transportation. However, estimating the localization of a user is not trivial, although it is critical for providing relevant information. I assess relevant design issues for a modular cost-efficient user-friendly on-trip Navigation service that uses position sensors. By helping travellers move from single-occupancy vehicles to public transportation systems, communities can reduce traffic congestion as well as its environmental impact. Here, I describe our efforts to increase the satisfaction of current public transportation users and help motivate more people to ride. I can help existing riders and encourage new riders by enhancing the usability of public transportation through good transit traveler information systems.*
*The motivation for every location based information system is: "To assist with the exact information, at right place in real time with personalized setup and location sensitiveness". LOCATION-BASED services are increasingly important for modern mobile devices such as the Smartphone.An important feature of a modern mobile device is that it can position itself. Not only for use on the device but also for remote applications that require tracking of the device. Furthermore, tracking has to robustly deliver position updates when faced with changing conditions such as delays due to positioning and communication, and changing positioning accuracy. The realized system tracks pedestrian targets equipped with GPS-enabled devices.*

## I. Introduction

Public transportation systems play an increasingly important role in the way people move around their communities. I consider some of the benefits of public transportation, the challenges facing its widespread adoption, and the role transit traveler information systems can play in meeting those challenges. For individuals, public transportation provides mobility to those who cannot or prefer not to drive, including access to jobs, education, and medical services. In general, transport mobility - the ability for people to move around their community - is a strong indicator for employment, with studies showing, for example, a direct connection between car ownership and employment. By helping travelers move from single-occupancy vehicles to public transportation systems, communities can reduce traffic congestion as well as its environmental impact.

Here, I describe our efforts to increase the satisfaction of current public transportation users and help motivate more people to ride. An important feature of a modern mobile device is that it can position itself. Not only for use on the device but also for remote applications that require tracking of the device. Furthermore, tracking has to robustly deliver position updates when faced with changing conditions such as delays due to positioning and communication, and changing positioning accuracy. The realized system tracks pedestrian targets equipped with GPS-enabled devices. I concentrate on the tools it provides for real-time arrival information, which is available through a variety of interfaces for mobile devices. Such information is valuable for both new and frequent riders. users could access information by navigating through a list of stops for a particular transit route. For the full Web interface, users could see stop and route information displayed on a map but still had to search for stops by stop number, route, or address.

Motivated by this consideration, I develop a location-aware native Smartphone application for BEST Bus that leverages the localization technology in modern mobile devices to quickly provide users with information for nearby stops and improved context-sensitive responses to their searches

## II. Literature Survey

The design of the navigation system was driven by a set of premises that distinguish it from other navigation solutions.
- The service should be deployable on short term, and not in a far future.
- Deployment cost for the service provider should be efficient.
- Usage cost should be low considering currently common communication costs.
- Service should be easily adaptable and extendable to a fast changing reality.

Next to deployability under current conditions, another concern had a major influence in system design: user friendliness. The system should require as little interaction from the user as possible and little change of his habits. Furthermore, all interfaces between user and navigation system should be carefully designed so that users unfamiliar with the system or with technology in general feel comfortable using it. Finally, being userfriendly also implies using technologies that most users are already familiar with, so that they do not have to acquire any new device or learn to handle one.

A navigation system complying to these design requirements enhances the experience of users of a PTN in a cost effective way. Visitors and sporadic users are target user groups, as the system is especially helpful for people unfamiliar with the PTN. But also normal users would profit, being guided to destinations out of their current and known parts of the transport network, for example visiting a place for the first time in an area they do not usually use. In this way, the navigation system enhances the urban mobility experience and makes using the PTN more attractive to people unfamiliar with it.

As with any application, it's important to consider the target audience. In this case, I can divide transit riders into new or infrequent riders, who aren't overly familiar with the local transit system, and frequent riders, who are familiar with it and use it every day. New or infrequent riders are less familiar with available routes and often need more trip-planning guidance, whereas frequent riders typically already know which sequence of stops and routes is the fastest to reach their destination, so they just want to know when the next bus is coming. The application presented in this article is targeted primarily at this second group of frequent transit users.

While static schedules and timetables are an important base for rider information, the reality is that transit vehicles do not always run on time. Traffic congestion, weather, accidents, and passenger incidents: there is any number of reasons why a transit vehicle might not meet its schedule. As such, many recent transit traveller information system improvements have focused on providing real-time arrival information.

## III. Proposed system

The system is composed of many pieces, tools, and interfaces:
- The website
- The Android app
- The API
- The service alerts webapp

**A. The Website**

The Website is the main entry point for most riders when using for the first time. The homepage has a description of the project, links to the various interfaces, and more details about the research driven by application.

There are three primary interfaces powered by the website:
- The standard desktop web interface;
- The Android-optimized mobile web interface; and
- The text-only web interface.

I will cover each interface briefly in the following subsections.

- **The Standard Desktop Web Interface**

The standard desktop web interface is designed to loosely mimic the interface of the main Google Maps website that many users are already familiar with. Specifically, the primary view is a Google map view, with a search field at the top and a search results panel on the left. Users can browse the map directly to see transit stops at a particular location, Additionally, users can search by route to display the map of that route and stops along the route. One important interface detail is that I calculate the direction of travel for routes serving a particular stop and show a directional arrow for the stop on the map. This direction of- travel arrow is particularly useful to riders when they are attempting to disambiguate between two transit stops that are right across the street from each other, but serve routes headed in opposite directions.

- **The Android-Optimized Mobile Web Interface**

The Android-optimized mobile web interface was designed first and foremost for the mobile browser on the Android. optimized webapps were the only outlet for running something like an "app" on the phone. These webapps basically consist of webpages optimized touch-screen web browser of the Android. When native apps were finally allowed on the Android, I eventually developed and released a native app for the Android that could take advantage of location-aware and mapping features of the phone. However, the Android-optimized web interface continued to be useful to users of other smart phones, such as the Blackberry, Android, which had a decent web browser but little-to-no native app support. That said, the graphics and layout of the Android-optimized mobile web interface are still too much for the simple text-only web browser on simple feature phones, which is why I have a text-only web interface as well, as described in the next section.

- **The Text-Only Web Interface**
The text-only web interface is designed to be usable with a very basic text-only web browser on a feature phone. It uses no graphics and very basic layouts to display information. The features offered and navigation are very similar to that of the Android-optimized mobile web interface, but just presented using a simpler layout.

**B. The Android App**
The native Android app provides a location-aware application for quickly accessing real-time arrival information for nearby public transit stops. Unlike the interfaces described thus far, the Android has built-in localization capabilities, using a fusion of sensor data from GPS, WiFi, and cell-tower localization to quickly get a location fix on a users phone. This location information can significantly reduce the time it takes to access real-time arrival information for a nearby stop. Beyond the key addition of the location-aware capabilities, the Andriod app has a lot of the same features available in the other interfaces: a map view, bookmarks, recent stop view, and search for stops by route, address, and stop number.

**C. The API**
The Android app is powered on the backend by web-based API, a standard technique for providing dynamic data to mobile apps. When I created the API for the Android app, I decided to make the API freely available to developers so that they might create apps of their own.

**D. The Service Alerts Webapp**
I believe that real-time public transit information of all types is important in improving the usability of public transit. One critical component of real-time information includes service alerts: details about reroutes, cancellations, delays, and other changes in service. Informing riders about these service alerts in a timely fashion is critical to improving the rider experience, but transit agencies currently struggle with this task. I believe that by providing tools to transit agencies that allow quick and easy management of service alert information that produces data in standardized forms, I can bypass the human bottle neck that is currently holding real-time service alerts back. Towards that end, I have created a service alerts webapp that can be used to quickly and easily manage service alert information. The webapp allows for the creation, modification, and deletion of complex service alerts

## IV. Concussion

I have presented my dissertation on the value of BEST Bus information systems, demonstrating a number of widely deployed tools and evaluations of those tools that show their utility. Specifically, I have described the system, which provides riders tools across a Smartphone and interfaces. I have demonstrated a real-time mobile trip planning tool and also a method for crowd-sourcing the detection of errors in public transit data. Finally, I have presented evaluations that show improves satisfaction with public transit, reduces wait times, increases transit usage, encourages walking, and improves perception of safety among riders.

Sustainable urban mobility is a key factor for a citizen's quality of life, as an increasingly larger amount of the population lives in urban areas. The integration and interoperability of different transport networks are seen in that document as a key feature for the improvement of urban mobility, together with improved travel information